\documentclass[review]{elsarticle}
\usepackage[T1]{fontenc}
\usepackage{lineno,hyperref}
\modulolinenumbers[5]
\usepackage{subcaption}
\usepackage{xcolor}
\usepackage{arydshln,amsmath}
\usepackage{soul}
\usepackage{comment}
\setstcolor{red}

\begin{document}

\begin{frontmatter}

\title{Study of excited electronic states of the $^{39}$KCs molecule correlated with the K($4^2$S)+Cs($5^2$D) asymptote: experiment and theory}

\author{Jacek Szczepkowski \corref{mycorrespondingauthor}}
\cortext[mycorrespondingauthor]{Corresponding authors}
\ead{jszczep@ifpan.edu.pl}
\author{Anna Grochola  \corref{c2}}
\author{Wlodzimierz Jastrzebski \corref{c1}}
\address{Institute of Physics, Polish Academy of Sciences, Al.  Lotnik\'ow 32/46,	02-668 Warszawa, Poland}

\author{Pawel Kowalczyk \corref{c3}}
\address{ Institute of Experimental Physics, Faculty of Physics, University of Warsaw, ul.~Pasteura~5, 02-093 Warszawa, Poland}

\author{Romain Vexiau}
\author{Nadia Bouloufa-Maafa}
\author{Olivier Dulieu}
\address{Universit\'e Paris-Saclay, CNRS, Laboratoire Aim\'e Cotton, B\^at. 505, rue du Belv\'ed\`ere, f-91400 Orsay, France}

\begin{abstract} 
Using the polarisation labelling spectroscopy, we performed the detailed analysis of the level structure of excited electronic states of the $^{39}$KCs molecule in the excitation energy interval between 17500~cm$^{-1}$ and 18600~cm$^{-1}$ above the $v=0$ level of the $X^1\Sigma^+$ ground state. We prove that the observed states are strongly coupled by spin-orbit interaction above 18200~cm$^{-1}$, as manifested by numerous perturbations in the recorded spectra. The spectra are interpreted with the guidance of accurate electronic structure calculations on KCs, including potential energy curves, transition electric dipole moments, and representation of the spin-orbit interaction with a quasi-diabatic effective Hamiltonian approach. The agreement between theory and experiment is found remarkable, clearly discriminating among the available theoretical data. This study confirms the accuracy of the polarisation labelling spectroscopy to analyse highly-excited electronic molecular states which present a dense level structure.
\end{abstract}
	
	\begin{keyword}
		 laser spectroscopy \sep KCs molecules \sep electronic states \sep
		potential energy curves \sep \emph{ab initio} calculations
		\PACS 31.50.Df \sep  33.20.Kf \sep 33.20.Vq \sep 42.62.Fi
	\end{keyword}

\end{frontmatter}

\section{Introduction}
\label{sec:intro}

For more than a century, diatomic molecules composed of two alkali-metal atoms are among the best-suited species to explore in great detail their structure and dynamics using the huge progress of both laser spectroscopy techniques and theoretical and numerical models. Indeed, their electronic transitions are easily accessible in the infrared to visible frequency domain, while their simple electronic structure with two valence electrons allows for very precise characterisation of their Hamiltonian. Thus, just like alkali-metal atoms have been at the source of spectacular developments of laser-cooling and quantum degenerate ultracold quantum gases, alkali-metal diatomic molecules have been the systems of choices to establish robust methods to create ultracold molecular gases, and to start exploring their potential impact in many areas, including ultra-high precision measurements and quantum information and simulation. This is particularly the case for the class of the so-called dipolar species composed of two different atoms, possessing a significant permanent electric dipole moment (PEDM) in their own frame.

The relative simplicity of such species allows many kinds of elaborate studies, like in the case of the present work, the exploration of the structure of their highly-excited electronic states. Their spectroscopic identification often becomes problematic as several electronic states could lie in the same energy range, sometimes interacting with each other, so that the assignment of the observed spectral lines could be ambiguous. Similarly, their theoretical representation involves a large configuration space, preventing them to be accurately  treated with conventional numerical approaches. Therefore accurate theoretical calculations on molecular states and their couplings are needed, and may provide invaluable guidance for unravelling intricate observations.

We present in this paper our combined spectroscopic and theoretical analysis of several excited electronic states of $^{39}$KCs molecule in the energy range close to the third excited K($4\,^2S_{1/2}$)+Cs($5\,^2D_{3/2,5/2}$) dissociation limit manifold. This work is a follow-up of our recent investigation of the molecular states correlated to the K($4\,^2P_{1/2,3/2}$)+Cs($6\,^2S_{1/2}$) dissociation limits \cite{SzczepkowskiKCs2018,SzczepkowskiKCs2020}, and complements an earlier study from the experimental group in Warsaw \cite{Szczepkowski2012}. The present excitation scheme is displayed in Fig.\ref{fig:experiment}. Starting from $^{39}$KCs molecules in their X$^1\Sigma^+$ ground state, we expect strong transitions to the (3)$^1\Pi$ state due to usual selection rules for dipole-allowed transitions. Instead, the recorded spectra reveal transitions to several electronic states, with lots of perturbations in the observed line series. These perturbations are presumably due to the spin-orbit interaction among the molecular states which is significant in the Cs($5\,^2D$) atom, and transferred to the KCs molecule.

In Section \ref{sec:experiment} we first present the principle of the experimental set up based on the powerful technique of polarisation labelling spectroscopy (PLS), and the details of the excitation scheme. We report in Section \ref{sec:theory_a} the electronic structure calculations relevant to this study, yielding results for potential energy curves (PECs), transition electric dipole moments (TEDMs), and molecular spin-orbit couplings (SOCs) depending on the internuclear distance $R$. In Section \ref{sec:theory_so} we develop our approach for including the spin-orbit interaction in the electronic structure calculations. The analysis of the recorded spectra is presented in Section \ref{sec:analysis}, relying on the identification of line progressions guided by the theoretical data.

\section{Experiment}
\label{sec:experiment}

In the experiment KCs molecules were formed in a stainless steel linear heat pipe oven as in previous experiments concerning this molecule \cite{Szczepkowski2012}. Approximately equal amounts (about 5~g each) of metallic potassium and caesium with a natural isotopic abundance were inserted to the central part of the heat pipe and heated to 390-430 degrees Celsius. Helium at pressure of around 5~Torr was used as a buffer gas to prevent interaction of the hot alkali metal vapour with the quartz windows.

The molecular spectra were recorded with the polarisation labelling spectroscopy (PLS) technique, which has been used for many years by our group for studies on alkali-metal diatomic molecules \cite{Szczepkowski2013}. This technique has been proved to be particularly relevant to interpret the spectra for highly-excited molecular states \cite{Jastrzebski2015}. PLS is a pump-probe experimental technique, in which simultaneous interaction of molecules with two properly-tuned and polarised laser beams allow to observe rovibrational molecular spectra only from a few chosen (“labelled”) levels of the ground state. Both laser beams cross at small angle in the molecular sample. The strong pump beam creates an optical anisotropy in the sample, which causes a change of polarisation of the second, weak probe beam, but only in the case when transitions induced by both beams originate from the same rovibrational level in the ground state. This change in polarisation is detected by a set of crossed polarisers. The key of the method is that the probe beam is tuned to a known transition in the investigated molecule which then selects ("labels") a specified ground state level. By tuning the pump laser across the investigated spectral range and recording changes of polarisation of the probe laser, we observe a greatly limited number of spectral lines with fully-resolved rotational structure. 

In this experiment an OPO/OPA laser system (Sunlite EX, Continuum)  pumped with third harmonic of an injection seeded Nd:YAG laser (Powerlite 8000) delivered the pump laser beam, which was scanned within the range of excitation energies $17500-18600$~cm$^{-1}$ above the minimum of the ground state potential. Two lasers were used for the probe laser: 
\begin{itemize}
    \item A ring laser working on Rhodamine 6G, with a wavelength controlled using a HighFinesse WS-7 wavemeter, and fixed on selected transitions from the X$^1\Sigma^{+}$ ground state to levels of the well known E(4)$^1\Sigma^{+}$ state~\cite{Szczepkowski2012};
    \item A home-made pulse dye laser working on Rhodamine 700 with a wavelength controlled in the same way. It was tuned to selected transitions between the X$^1\Sigma^{+}$ ground state and the B$^1\Pi$ state \cite{Birzniece2012}.
\end{itemize}

The rotationally resolved excitation spectra of KCs were recorded with accuracy of line positions better than 0.1~cm$^{-1}$. The calibration of molecular spectra was achieved by simultaneous recording of argon and neon optogalvanic spectra and transmission fringes from a Fabry-P\'{e}rot interferometer with FSR~=~1~cm$^{-1}$.

\begin{figure}
	\includegraphics[width=0.95\linewidth]{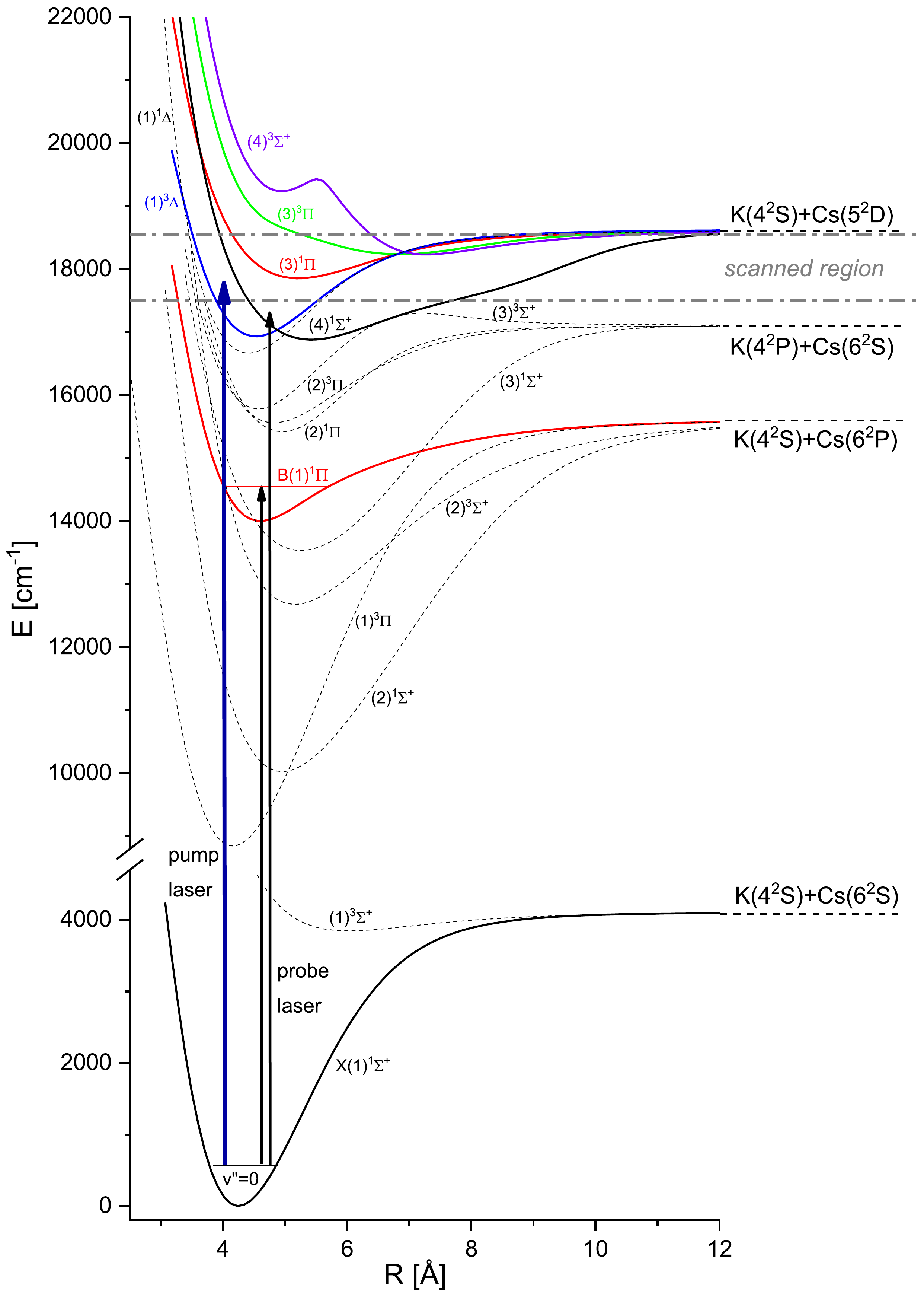}
	\caption{The KCs electronic transitions involved in the present PLS experiment, with the corresponding PECs extracted from our calculations described in Section \ref{sec:theory_a}. The strong pump laser excites molecules from the X$^1\Sigma^{+}$ ground state to one of the states correlating to the K($4S$)+Cs($5D$) asymptote. The weak probe laser excites molecules from the X$^1\Sigma^{+}$ ground state to one of the two previously characterised electronic states B$^1\Pi$ or E(4)$^1\Sigma^{+}$.}
	\label{fig:experiment}
\end{figure}

The choice of two possibilities for the probe laser enhances the intensities of lines corresponding to transitions to $e$-parity (resp. $f$-parity) levels when it is tuned to the E $\leftarrow$ X (resp. B $\leftarrow$ X) transitions \cite{Ferber199753}. All the relevant transitions are summarised in Fig.\ref{fig:experiment} where the corresponding PECs are highlighted in colour (see the online version).

\section{Electronic structure calculations of Hund's case (a) potential energy curves}
\label{sec:theory_a}

In an earlier work \cite{Vexiau2017}, we have calculated the Hund's case (a) PECs of all heteronuclear alkali-metal diatomic molecules (including KCs) to study the dynamical dipolar polarisabilities of their ground state with the aim of finding the best laser frequency allowing their optimal optical trapping. We used these data for the lowest excited electronic states to make prospects for the formation of ultracold polar ground state KCs molecules via an optical process \cite{Borsalino2016}. Furthermore, we modelled the fine and hyperfine structure of the electronic states dissociating to the second and third dissociation limits K(4$^2$S)+Cs(6$^2$P$_{1/2,3/2}$) and K(4$^2$P$_{1/2,3/2}$)+Cs(6$^2$S), respectively  \cite{Orban2015,Orban_KCs_HFs2019}. We noted that our results for such low-lying electronic states were in good agreement with the PECs reported in \cite{Korek2000,Kim2009}. Our calculations rely on an approach extensively described in \cite{aymar2005,aymar2006a,guerout2010}, that we briefly recall in the following.

Computing PECs for highly-excited electronic molecular states is usually a challenging task, as the structure of the corresponding electronic wavefunctions could be spread over many Slater determinants generated in the considered configuration space. Therefore, implementing a full configuration interaction (FCI) in the Hilbert space generated by a large Gaussian basis set yields a good chance for obtaining reliable PECs. The KCs molecule is described as an effective two-valence electron molecule moving around K$^+$ and Cs$^+$ polarisable cores. The ionic cores are described by appropriate effective core potentials (ECP) (see \cite{aymar2005,aymar2006a}, and references therein), with additional parametric core polarisation potentials (CPPs) (see \cite{guerout2010}, and references therein) to take into account the correlation between core and valence electrons in an effective manner. The electronic Hamiltonian matrix for the two-electron system is expressed in a large Gaussian basis set for every Hund's case (a) molecular symmetry, and fully diagonalised (namely, the FCI), yielding the requested PECs for a large number of electronic states. Permanent and transition dipole moments (PEDMs and TEDMs, respectively) are also extracted as functions of the internuclear distance $R$.

\begin{figure}
	\includegraphics[width=0.95\linewidth]{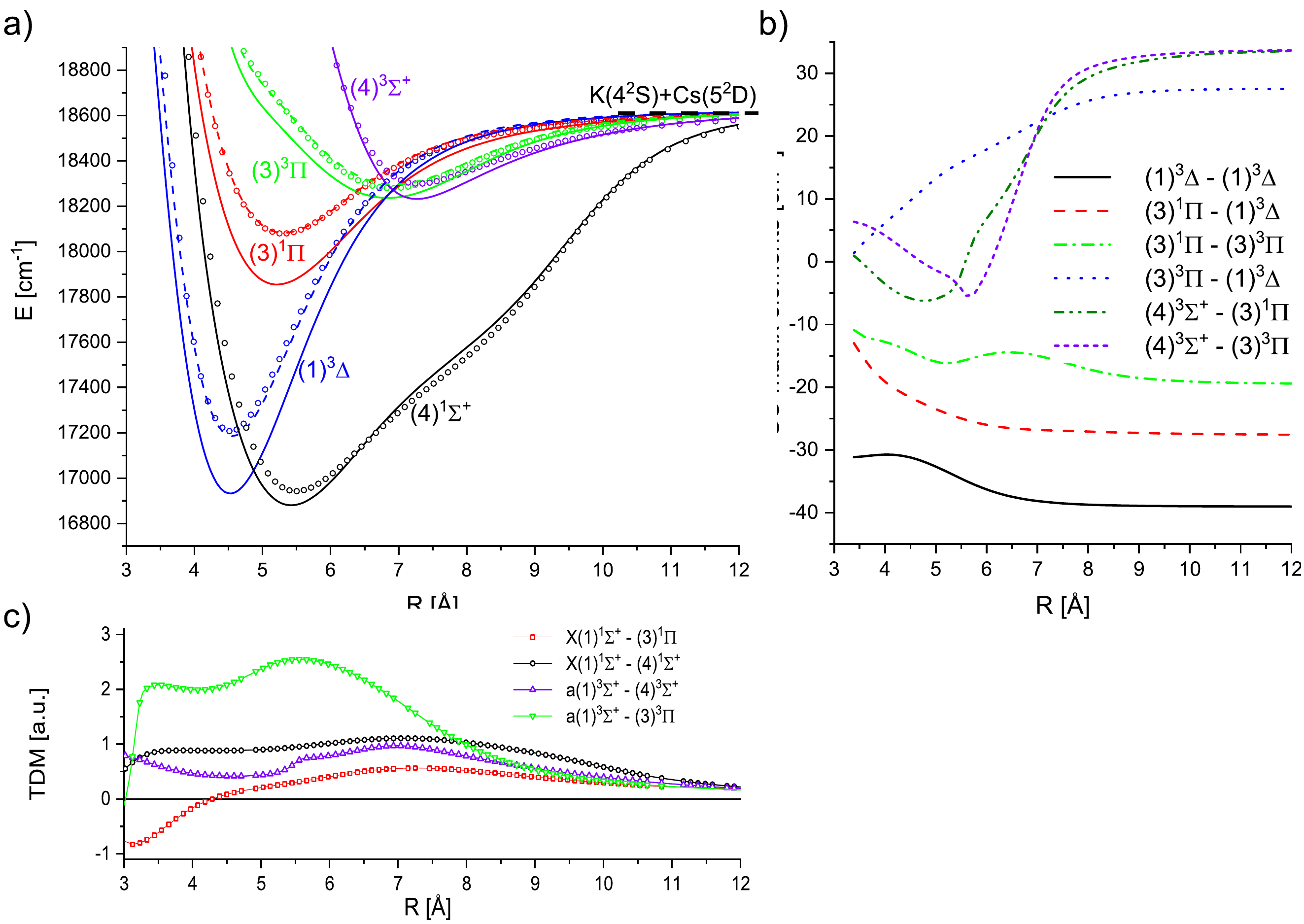}
	\caption{(a) Hund's case (a) potential energy curves of the electronic states correlating to the K($4^2S$)+Cs($5^2D$) asymptote, computed with the present approach (solid lines), and from Ref.~\cite{Korek2000} (open symbols), and Ref.~\cite{Habli2020} (dashed lines).  (b) Spin-orbit coupling matrix elements for the subspace of states with projection of the total electronic angular momentum $\Omega=1$. (c) Transition electronic dipole moments from the X$^1\Sigma^{+}$ and a$^3\Sigma^{+}$ state to the states of panel (a) allowed by selection rules.}
	\label{fig:pot_Hund_a}
\end{figure}

The PECs and TEDMs relevant for the present spectral region are displayed in Fig.\ref{fig:pot_Hund_a}, and PECs are compared to those of Refs. \cite{Korek2000,Habli2020}. Significant differences are found which at first sight is somewhat surprising as the methodology used in these references is very close to ours. The semi-empirical parameters involved in the definition of the ECPs and CPPs, namely the atomic ion static dipole polarisability and the cut-off radii, are basically equivalent. The main difference consists in the choice of the Gaussian basis set, which does not contain $f$ orbitals in Refs.\cite{Korek2000,Habli2020}, while our calculations include two such orbitals for K atom and four for Cs atom \cite{aymar2005}.

\begin{figure}
	\includegraphics[width=0.95\linewidth]{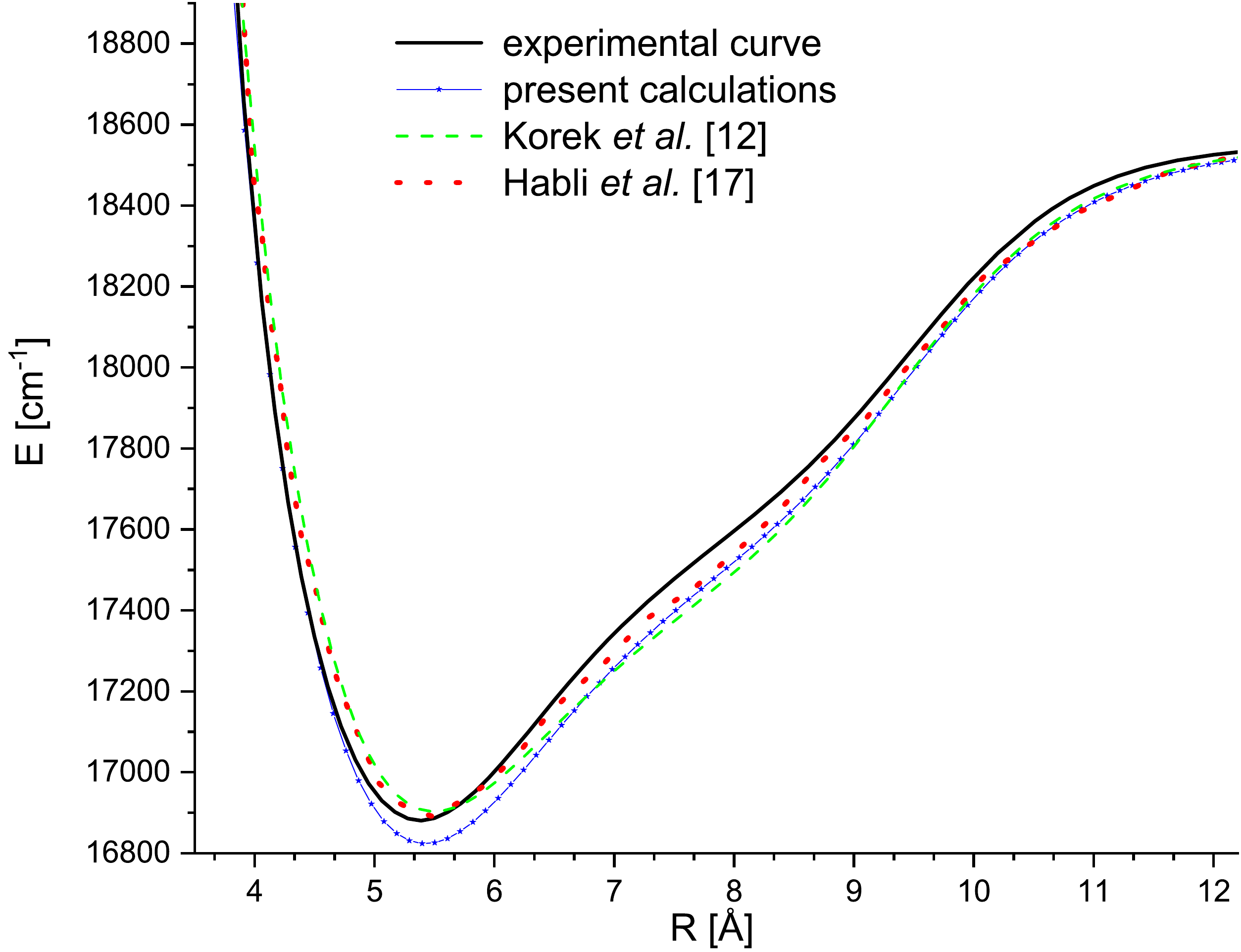}
	\caption{Comparison of the experimental PEC of the (4)$^1\Sigma^+$ state previously obtained in~\cite{Szczepkowski2012} (solid black line) with the present calculations (blue line with  circles), Korek~\textit{et~al.}~\cite{Korek2000} (green dashed line) and Habli~\textit{et~al.}~\cite{Habli2020} (red doted line).}
	\label{fig:4sigma}
\end{figure}

For completeness, we display in Fig.\ref{fig:4sigma} the comparison of the PEC of the (4)$^1\Sigma^+$ state experimentally determined in \cite{Szczepkowski2012}, with the  present Hund's case (a) curve, showing a satisfying agreement especially regarding the overall shape of the PEC.

\section{Quasidiabatic effective Hamiltonian for the spin-orbit interaction}
\label{sec:theory_so}

The expression of the spin-orbit (SO) operator for the molecule may be tedious to establish, especially when effective core potentials are used so that an effective expression must be developed. Here we chose a different approach, based on the definition of an effective \textit{quasidiabatic} Hamiltonian, as it was first proposed in Ref. \cite{cimiraglia1985}, and used for computing the SO coupling in the NaCd van der Waals molecule \cite{angeli1996}. The procedure is applied for every Hund's case (a) symmetry $(^{2S+1}\Lambda^{p})$,  where $S$, $\Lambda$, $p$ are the quantum numbers for the total electronic spin, the projection of the total electronic angular momentum on the molecular axis, and the parity for reflection with respect to a plane containing the molecular axis. In the present case, we consider the PECs with symmetries $^1\Sigma^+$, $^3\Sigma^+$, $^1\Pi$, $^3\Pi$, $^1\Delta$, $^3\Delta$, correlated to the asymptotes up to K($4^2S$)+Cs($5^2D$), see Fig.\ref{fig:experiment}.

Following the standard Born-Oppenheimer approximation, we first calculate the $N$ lowest eigenvalues \{$V_j(R)$\} and electronic eigenfunctions \{$|\Psi_j^0 (R)>$\} ($j=1,...,N$) of the electronic Hamiltonian $\hat{\textbf{H}}^0$, depending parametrically on the internuclear distance $R$, for every Hund's case (a)  molecular symmetry $^{2S+1}\Lambda^{p}$. These energies constitute a $N \times N$ $R$-dependent diagonal matrix $\textbf{H}^0(^{2S+1}\Lambda^{p};R)$ at every distance $R$. They represent the Hund's case (a) PECs.

At a sufficiently large internuclear distance $R_{ref}$, we build up a basis of $N$ ''reference states'' \{$|\textbf{R}_j>$\} $\equiv$ \{$|\Psi_j^0 (R{ref})>$\} ($j=1,...,N$) which describes the separated atoms with a good accuracy, for every $^{2S+1}\Lambda^{p}$ symmetry.

The third step aims at obtaining, at every internuclear distance $R$, a unitary \textbf{T}-matrix which connects the adiabatic matrix $\textbf{H}^0(^{2S+1}\Lambda^{p};R)$ to a quasi-diabatic matrix $\textbf{H}_{diab}(^{2S+1}\Lambda^{p};R)=\textbf{TH}^0(^{2S+1}\Lambda^{p};R)\textbf{T}^T$: the objective is to determine and optimise \textbf{T} in order to generate a new basis set $|\Psi^0>\textbf{T}$ which is as similar as possible to the reference basis set $|\textbf{R}>$. If completed, the adiabatic Hamiltonian is now expressed in an effective basis representing the two separated atoms.

\begin{figure}
\includegraphics[width=0.7\textwidth]{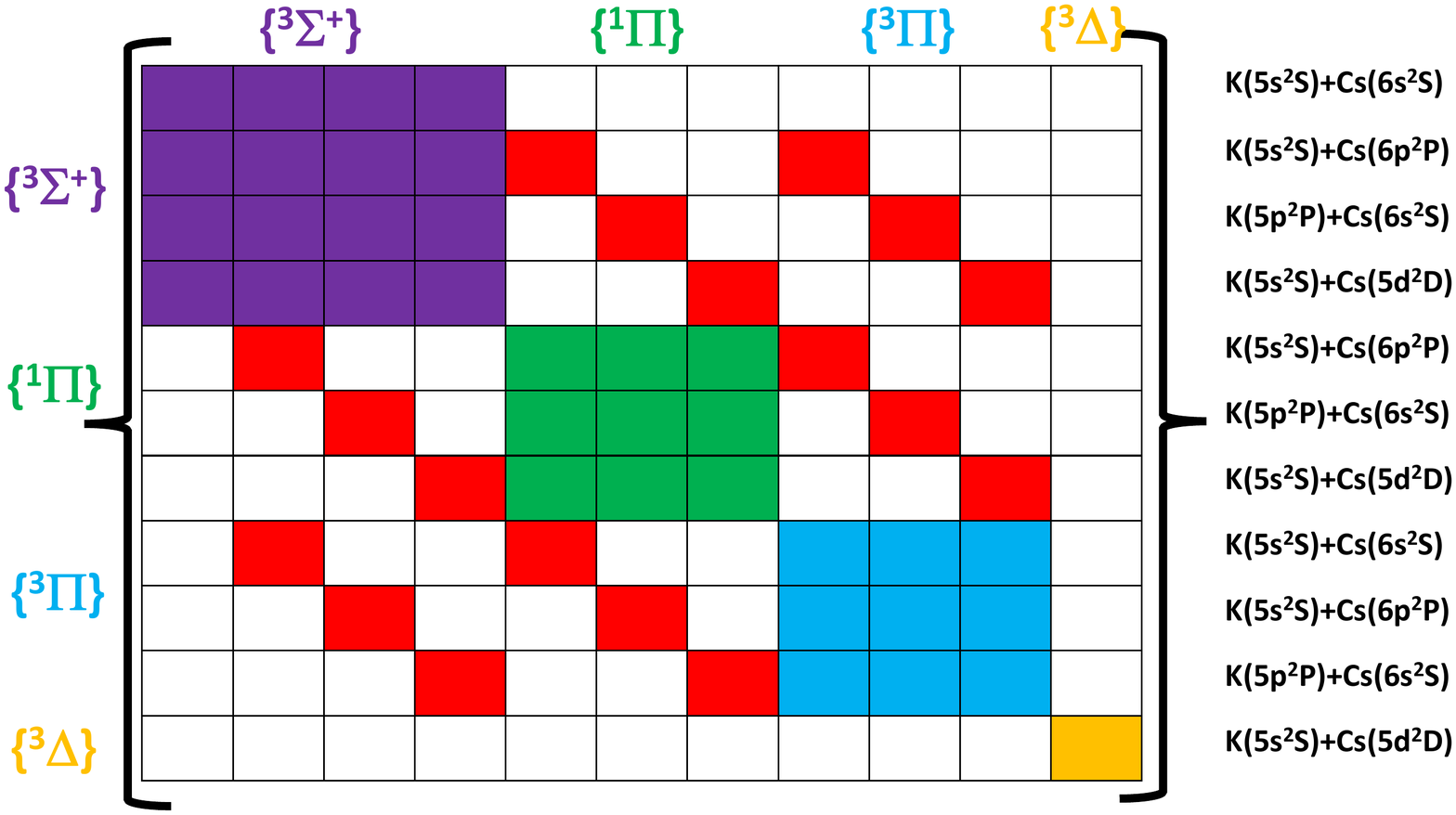}
\caption{Schematic view of the effective Hamiltonian matrix $\textbf{H}_{eff}$, illustrated in the case of $\Omega=1$. Four diagonal blocks are present for the four relevant Hund's case (a) symmetries $^3\Sigma^+$ (purple), $^1\Pi$ (green), $^3\Pi$ (blue), $^3\Delta$ (orange) contributing to the $\Omega=1$ manifold. The atomic SOCs (red cells, see \ref{eq:atomicsoc}) connect states from different blocks correlated to the same dissociation limit (thus the same atomic fine structure manifold). The remaining blank cells contain zeros.}
\label{fig:Heff}
\end{figure}

The spin-orbit matrices $\textbf{H}_{so}(\Omega^p)$, coupling the relevant Hund's case (a) states for every Hund's case (c) symmetry $\Omega^p$, are known in the separated atom representation: they involve energies of atomic fine structure splitting at each dissociation limit, and are thus independent of $R$. Therefore an effective Hamiltonian matrix including spin-orbit interaction, $\textbf{H}_{eff}(\Omega^p;R)=[\textbf{H}_{diab}(^{2S+1}\Lambda^{p};R)]+\textbf{H}_{so}(\Omega^p)$, is built by putting together the quasidiabatic matrices for the coupled molecular symmetries $^{2S+1}\Lambda^{p}$ (symbolised by the brackets) as diagonal blocks, and distributing diagonal and off-diagonal terms between blocks, according to the expression for $\textbf{H}_{so}(\Omega^p)$. A schematic representation of $\textbf{H}_{eff}(\Omega^p;R)$ is displayed in Fig.~\ref{fig:Heff}.

The diagonalisation of $\textbf{H}_{eff}(\Omega^p;R)$ yields the adiabatic Hund's case (c) PECs for every $\Omega^p$ symmetry. Alternatively, the inverse transformation of $\textbf{H}_{eff}(\Omega^p;R)$ back to the original adiabatic basis $|\Psi^0>$ leads to a new  matrix $\mathcal{H}_0(R)(\Omega^p;R)$: we extract the $R$-dependent diagonal and off-diagonal matrix elements for the molecular spin-orbit couplings between the involved molecular symmetries by the difference between the terms of $\mathcal{H}_0(R)$ and those of $\textbf{H}^0$ for all relevant $(^{2S+1}\Lambda^{p})$ symmetries. These coupling terms are displayed in \ref{fig:pot_Hund_a}b. Their asymptotic values match the values of the atomic spin-orbit coupling matrix

\begin{equation}
\begin{pmatrix}
 & ^3\Sigma^+ & ^1\Pi & ^3\Pi & ^3\Delta \\ 
 \hline
  ^3\Sigma^+ & 0 & \dfrac{A\sqrt{3}}{2} & \dfrac{A\sqrt{3}}{2} & 0 \\
   ^1\Pi & \dfrac{A\sqrt{3}}{2} & 0 & -\dfrac{A}{2} & -\dfrac{A\sqrt{2}}{2} \\
  ^3\Pi & \dfrac{A\sqrt{3}}{2} & -\dfrac{A}{2} & 0 & \dfrac{A\sqrt{2}}{2} \\
  ^3\Delta & 0 & -\dfrac{A\sqrt{2}}{2} & \dfrac{A\sqrt{2}}{2} & -A \\
\end{pmatrix},
\label{eq:atomicsoc}
\end{equation}

\noindent where $A=2 \Delta E_{fs}/5$ is the strength constant related to the atomic fine structure splitting $\Delta E_{fs}=97.58552$~cm$^{-1}$ of the Cs($5\,^2D$) level \cite{NIST_ASD}.

\begin{figure}
	\includegraphics[width=0.95\linewidth]{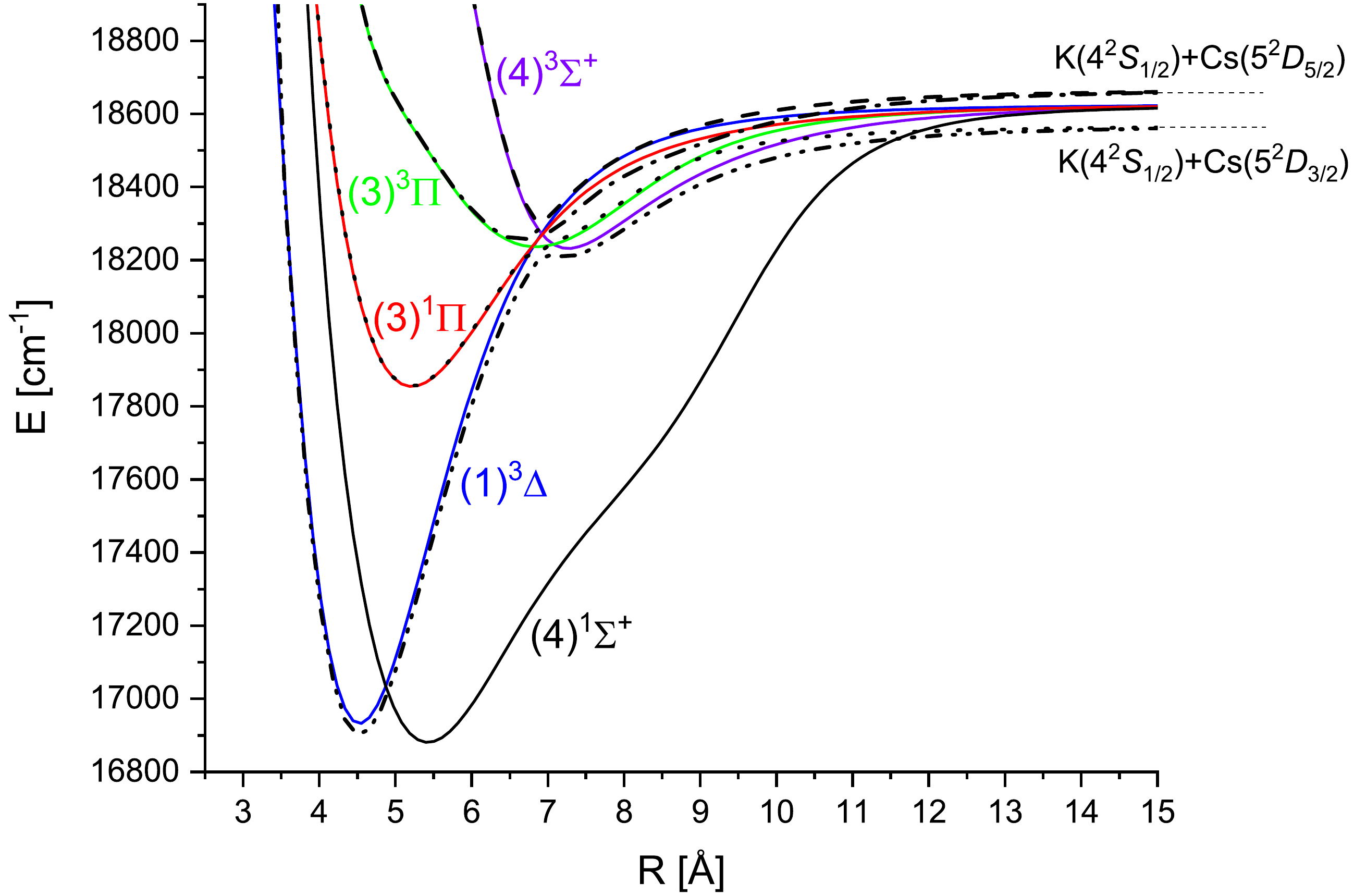}
	\caption{Calculated potential energy curves for Hund's case (a) converging to the K(4$^2$S)+Cs(5$^2$D)  asymptote (solid lines), and for Hund's case (c) with $\Omega=1$, dissociating to the K(4$^2$S$_{1/2}$)+Cs(5$^2$D$_{3/2,5/2}$)  asymptotes separated by $\Delta E_{fs}=97.58552$~cm$^{-1}$ (dashed lines).}
	\label{fig:pot_Hund_ac}
\end{figure}

In Fig.\ref{fig:pot_Hund_ac} we compare the Hund's case (a) and case (c) PECs calculated with the present approach, in the case of $\Omega=1$. As expected the main effect of the SOC is visible at the location of the crossing between the four PECs of $^3\Sigma^+$, $^1\Pi$, $^3\Pi$, $^3\Delta$ symmetries around 7\AA, resulting into several avoided crossings in Hund's (c) PECs, which should generate perturbations in the molecular spectra. At shorted distances, the PECs are shifted by a fraction of the atomic fine structure splitting as suggested by the molecular SOCs of Fig.\ref{fig:pot_Hund_a}.

\begin{figure}
	\includegraphics[width=0.95\linewidth]{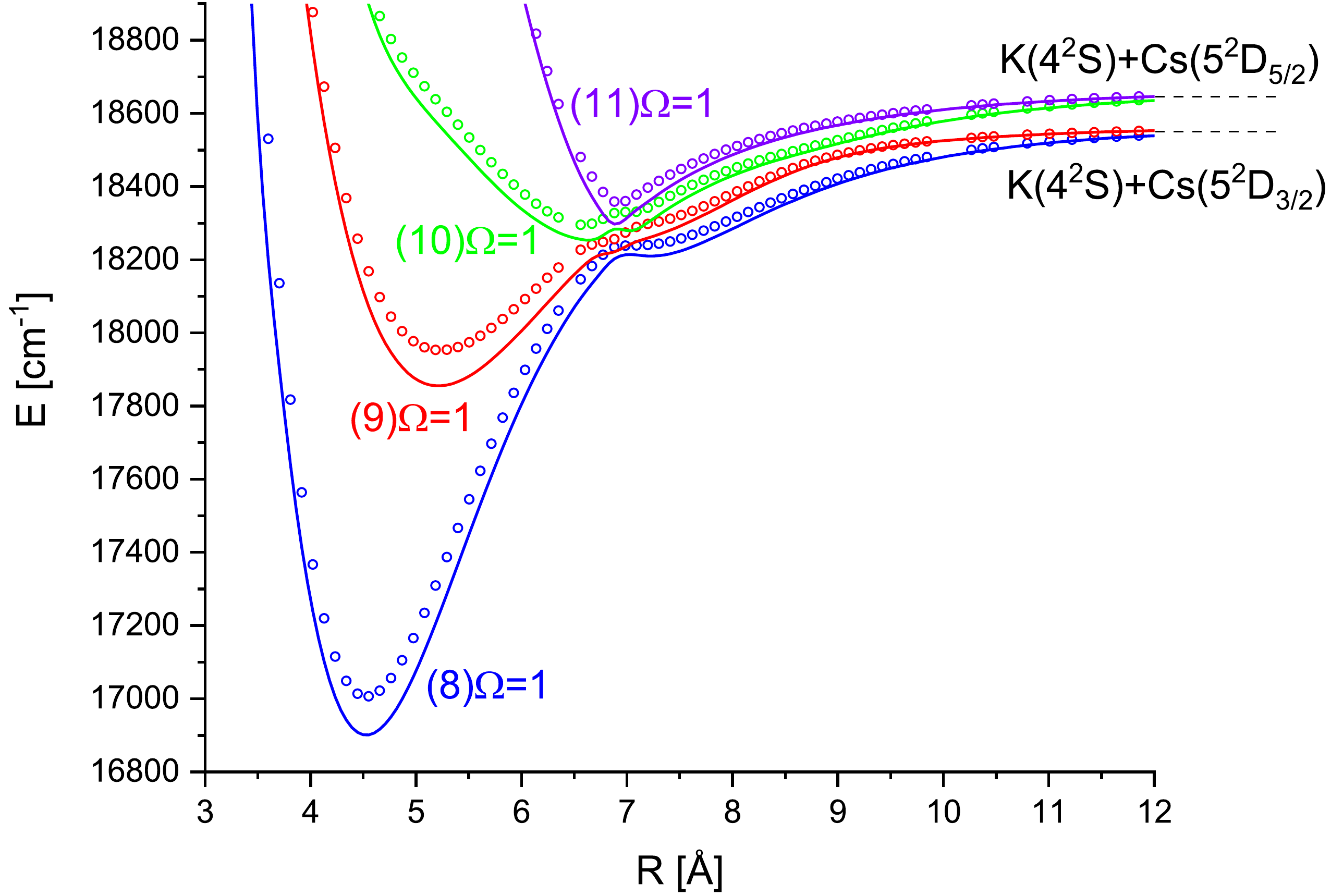}
	\caption{Potential energy curves of the four electronic states correlated to the K(4$^2$S$_{1/2}$)+Cs(5$^2$D$_{3/2,5/2}$) asymptotes (Hund's coupling case (c)), for which energies of rovibrational levels were found experimentally. Calculations by Korek \emph{et al.}~\cite{Korek2006} are represented by points, results of the present work by solid lines.}
	\label{fig:pot_Hund_c}
\end{figure}

Korek \emph{et al.}~\cite{Korek2006} also provided PECs including diagonalised molecular spin-orbit operator, and we compare them to our results in Fig.\ref{fig:pot_Hund_c}. Both calculations are in good agreement on the right-hand side of the avoided crossings, while the differences at shorter distances mainly reflect those already invoked for Hund's case (a) PECs in Fig. \ref{fig:pot_Hund_a}. Unfortunately the SOC matrix elements cannot be directly compared as they were not reported in \cite{Korek2006}.

Numerical data described in this Section can be found, in the online version~\cite{supplC}

\section{Analysis of the recorded PLS spectra}
\label{sec:analysis}

In the course of the experiment around 2400 spectral lines belonging to the $^{39}$KCs molecule were recorded in the spectral region $17500-18600$~cm$^{-1}$. No lines corresponding to other isotopes were observed, as they were much weaker and overlapped by strong lines belonging to the main isotope. The observed spectral lines correspond to transitions from the ground state of KCs to several excited electronic states correlated to the asymptotes K(4$^2$S$_{1/2}$)+Cs(5$^2$D$_{3/2,5/2}$). In the Hund's case (a) framework six electronic states correlate to the K(4$^2$S)+Cs(5$^2$D) asymptote, namely (4)$^1\Sigma^{+}$, (3)$^1\Pi$, (4)$^3\Sigma^{+}$, (3)$^3\Pi$, (1)$^1\Delta$, and (1)$^3\Delta$. At the first stage, P and R lines assigned to transitions to the known high vibrational levels of the (4)$^1\Sigma^{+}$ state ($v’$>50) \cite{Szczepkowski2012} were eliminated from the data set. 

\begin{figure}
	\includegraphics[width=0.95\linewidth]{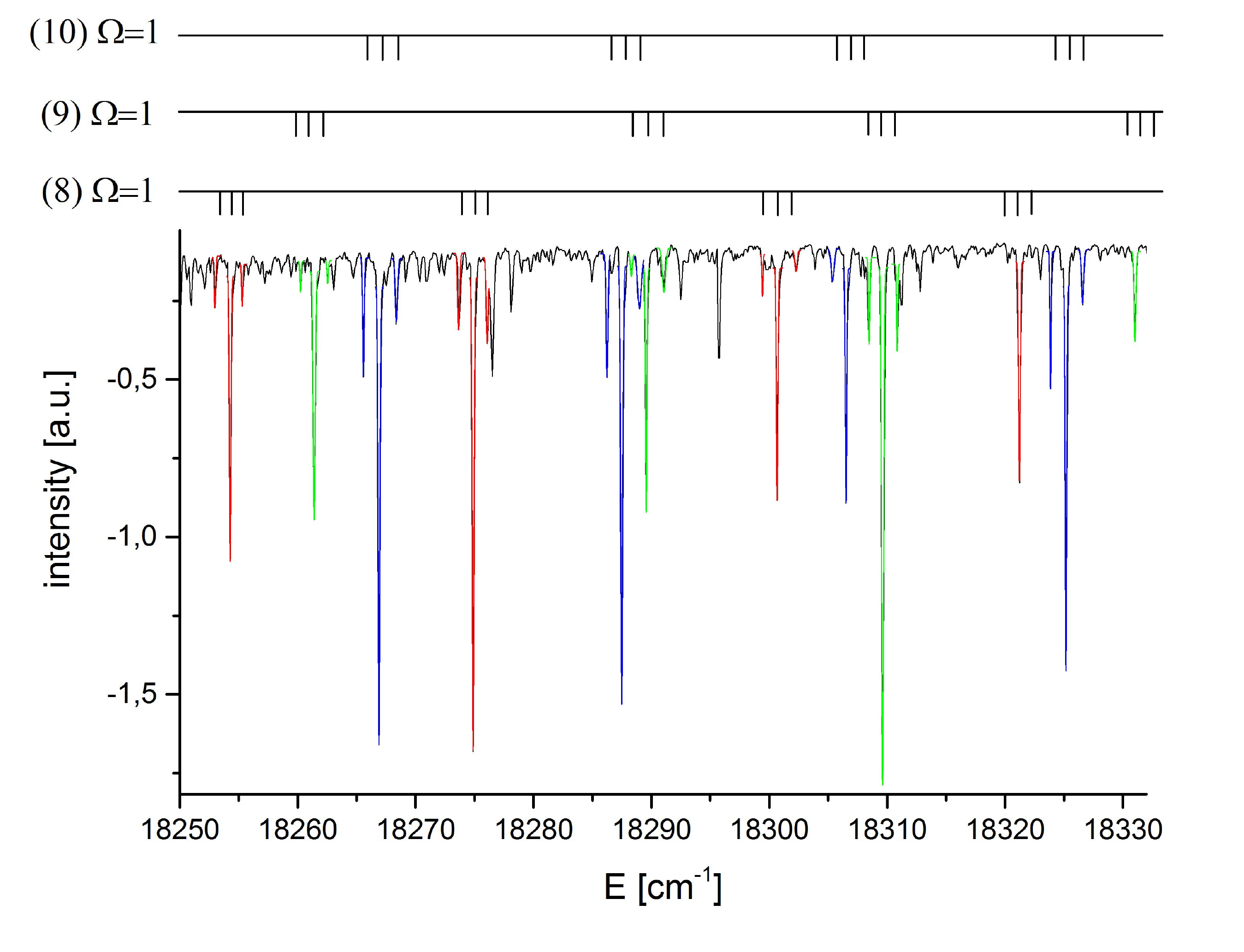}
	\caption{A portion of experimental spectrum of $^{39}$KCs with spectral lines corresponding to transitions from the rovibrational level $v''=0$, $J''=46$ of the X$^1\Sigma^{+}$ ($\Omega$=$0^{+}$) electronic ground state to consecutive rovibrational levels of three excited electronic states of $\Omega = 1$ symmetry (marked with red, green and blue in colour).}
	\label{fig:spectrum}
\end{figure}

This revealed groups of P, Q and R lines as shown in Fig.\ref{fig:spectrum}. In Hund's case (a) description this would suggest transitions from the $^1\Sigma^+$ ground state to a $^1\Pi$ excited electronic state. However, as transitions to more than one electronic state were undoubtedly recorded (marked in Fig.\ref{fig:spectrum} with three different colours), we deduced that a simple Hund's case (a) description is inappropriate and Hund's case (c) should be used instead (which might have been expected as $^{39}$KCs is a relatively heavy molecule).

\begin{figure}
	\includegraphics[width=0.95\linewidth]{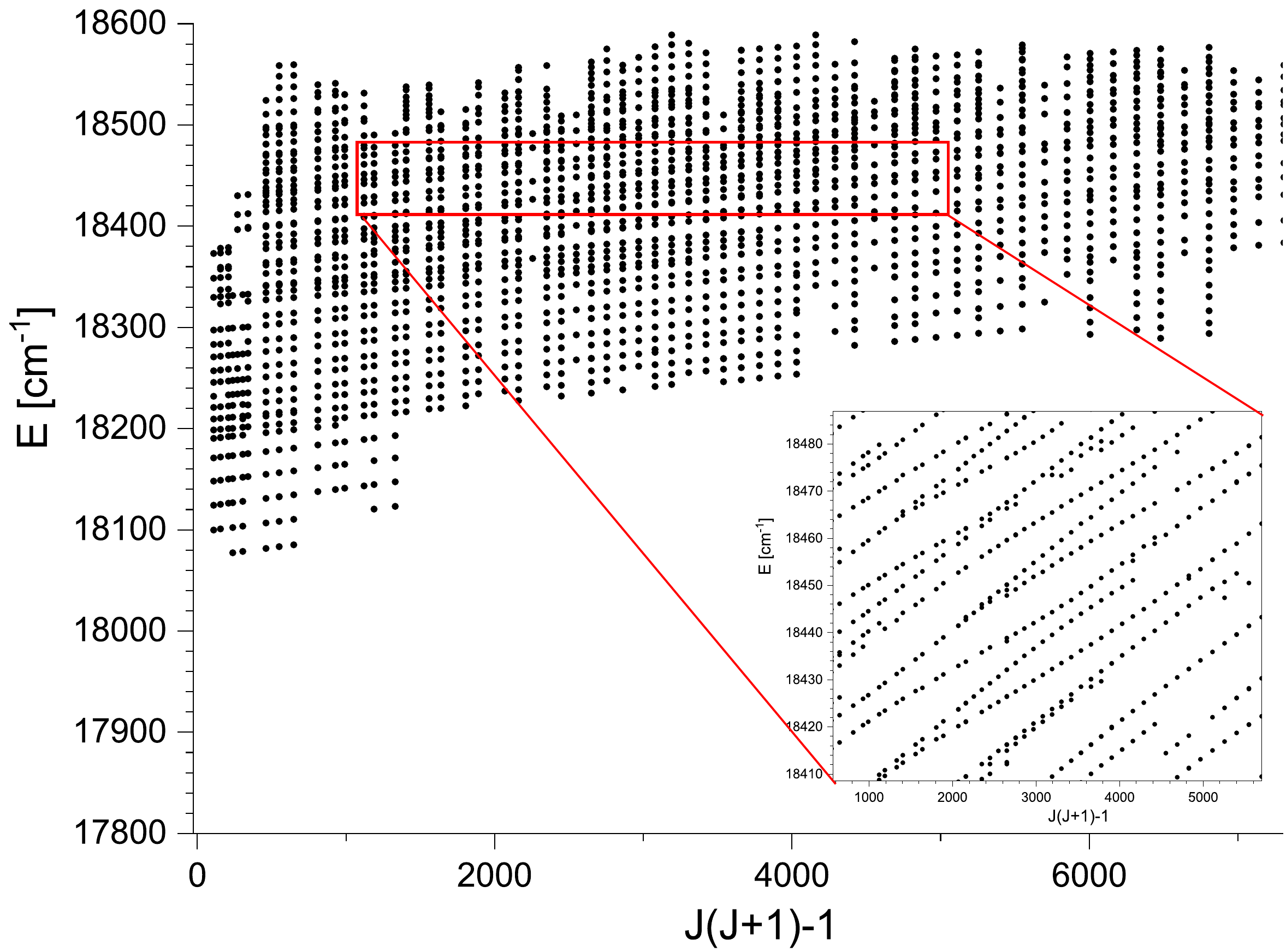}
	\caption{Experimental energies of the recorded rovibrational levels belonging to a mixture of (4)$^1\Pi$, (4)$^3\Sigma^{+}$, (4)$^3\Pi$, and (1)$^3\Delta$ electronic states plotted as function of $J(J+1)-1$ values, where $J$ is the rotational quantum number of levels of the excited electronic states.}
	\label{fig:exp_data}
\end{figure}

In Hund's case (c) picture the observed spectral lines can correspond to transitions from levels of the X$^1\Sigma^{+}$($\Omega$=$0^{+}$) ground state to levels of four $\Omega = 1$ excited electronic states resulting from (3)$^1\Pi$, (4)$^3\Sigma^{+}$, (3)$^3\Pi$, and (1)$^3\Delta$ states. Thus one may expect that transitions to four states, labelled (8)$\Omega$=1, (9)$\Omega$=1, (10)$\Omega$=1, and (11)$\Omega$=1, were recorded. Indeed, as can be seen in Fig. \ref{fig:exp_data}, in which term values of the observed levels are plotted against $J(J+1)-1$, and distribution of the experimental data is displayed, the observed rovibrational levels belong to more than one electronic state (see particularly the inset of the Figure). 

However in pure Hund's case (c) the theory predicts that transitions to the deepest of the observed states, the (8)$\Omega$=1, should be observed almost throughout the whole range of its vibrational levels, but experimentally they appear only in approximately the same energy range as transitions to the next (9)$\Omega$=1 state. Such observation indicates that none of the pure Hund's coupling cases can be used to describe the investigated states. Spin-orbit coupling between the states is apparently strong enough to record forbidden transitions to states of triplet symmetry, even doubly-forbidden transitions to the (1)$^3\Delta$ state, but only in the region where transitions to the (3)$^1\Pi$ state can be also observed. In other words, all the observed levels should belong to excited electronic states which contain a noticeable fraction of the (3)$^1\Pi$ state to ensure non-vanishing transitions probability.

In the lower part of the investigated energy region, in which transitions to only two electronic states are visible, (3)$^1\Pi$ and (1)$^3\Delta$, separate vibrational levels belonging to particular electronic states can be discerned. Unfortunately, we were not able to observe Franck-Condon oscillations in the intensities of the recorded spectral lines, thus determination of the absolute vibrational numbering of energy levels is impossible. However, guided by theoretical predictions, we can limit the range of possible vibrational quantum numbers for these levels.

\begin{figure}
    \centering

      \begin{subfigure}[t]{0.47\textwidth}
    \caption{ Hund's case (a)~\cite{Habli2020}}
        \includegraphics[width=\textwidth]{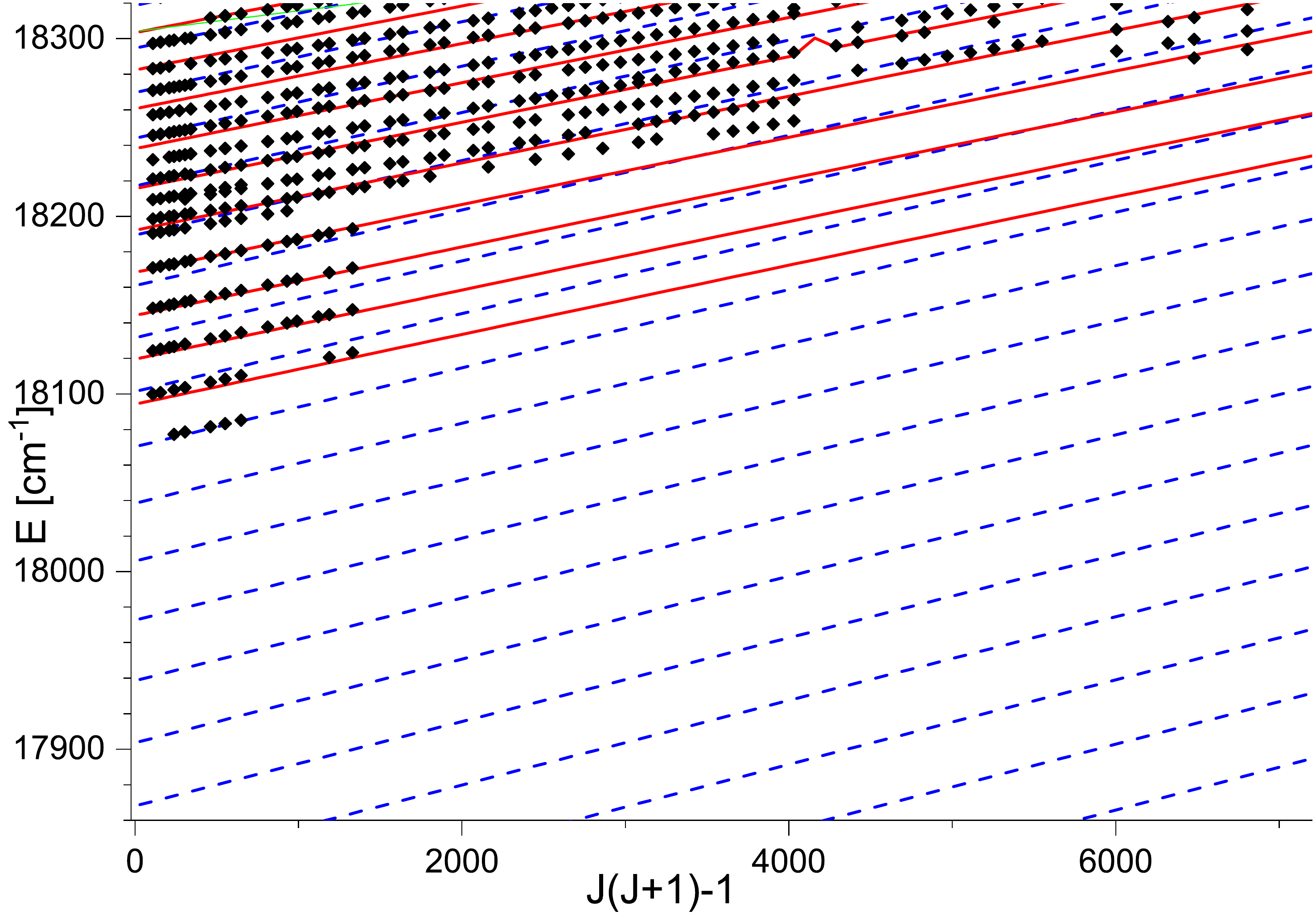}
        \label{fig:HabliA}
    \end{subfigure}
    \hfill
  \begin{subfigure}[t]{0.45\textwidth}
    \end{subfigure}

    \begin{subfigure}[t]{0.47\textwidth}
    \caption{ Hund's case (a)~\cite{Korek2000}}
        \includegraphics[width=\textwidth]{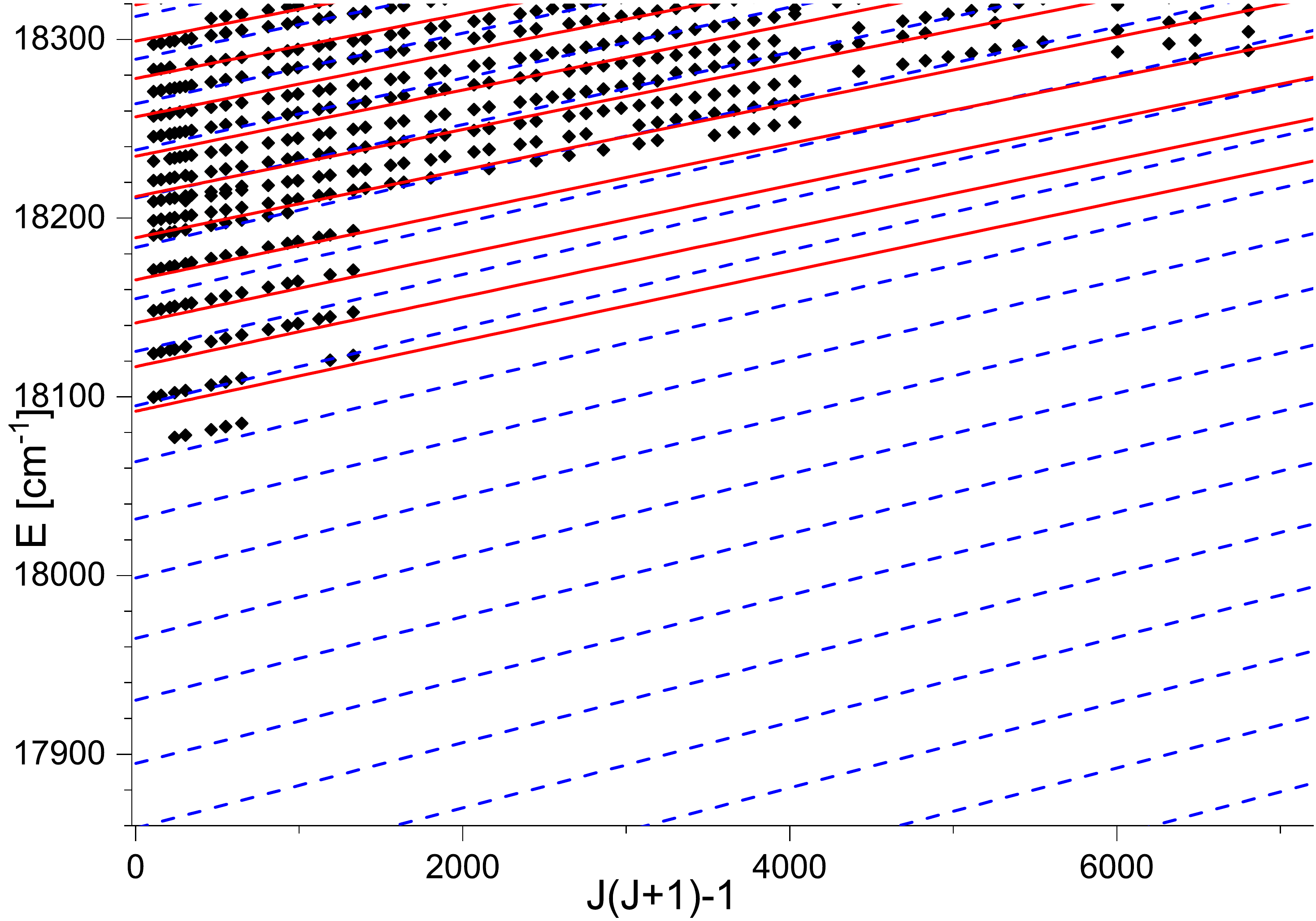}
        \label{fig:KorekA}
    \end{subfigure}
    ~ 
    \begin{subfigure}[t]{0.47\textwidth}
\caption{Hund's case (a)~PC}
        \includegraphics[width=\textwidth]{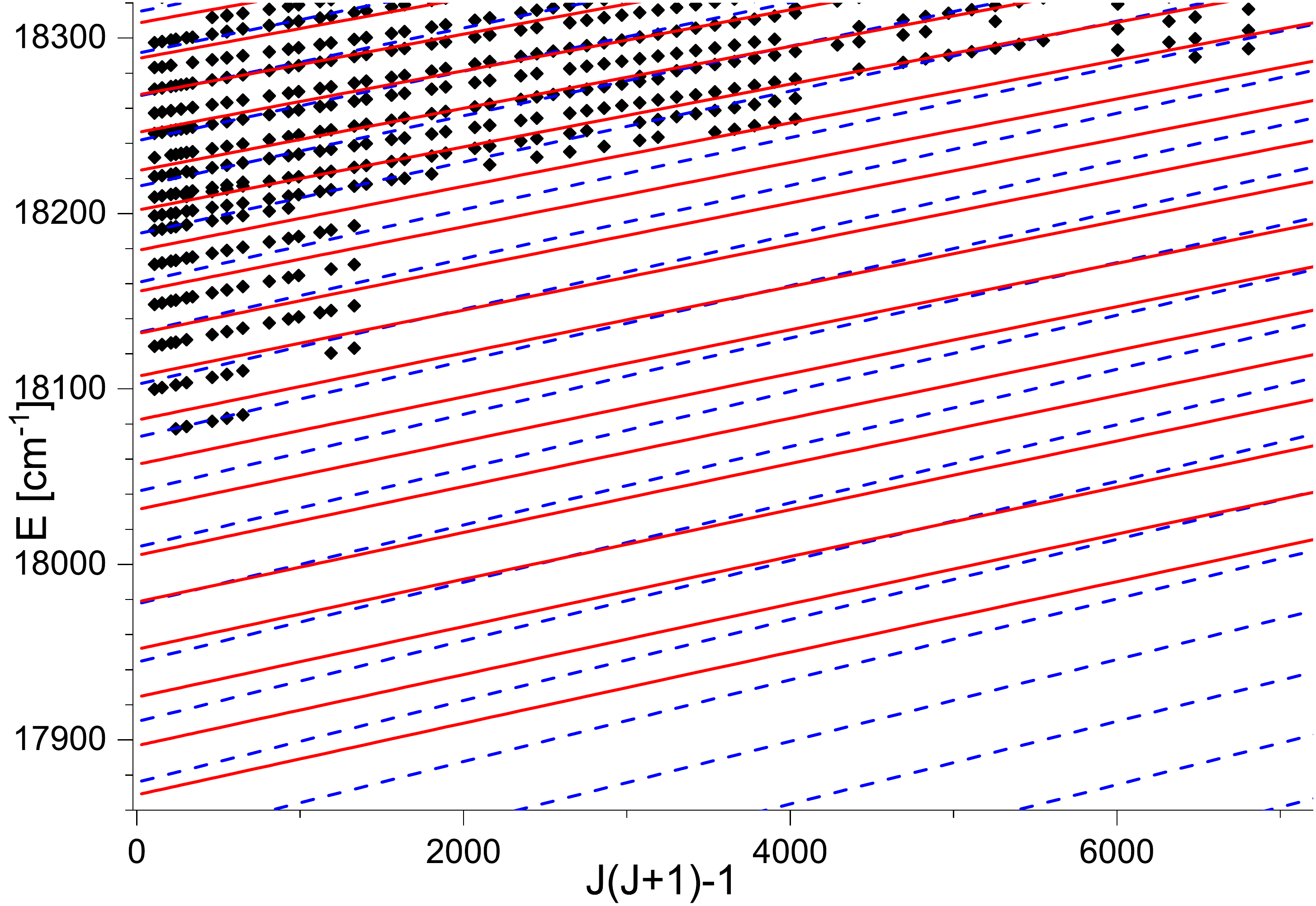}
        \label{fig:OlivierA}    
    \end{subfigure}
    ~
    
        \begin{subfigure}[t]{0.47\textwidth}
    \caption{Hund's case (c)~\cite{Korek2006}}
        \includegraphics[width=\textwidth]{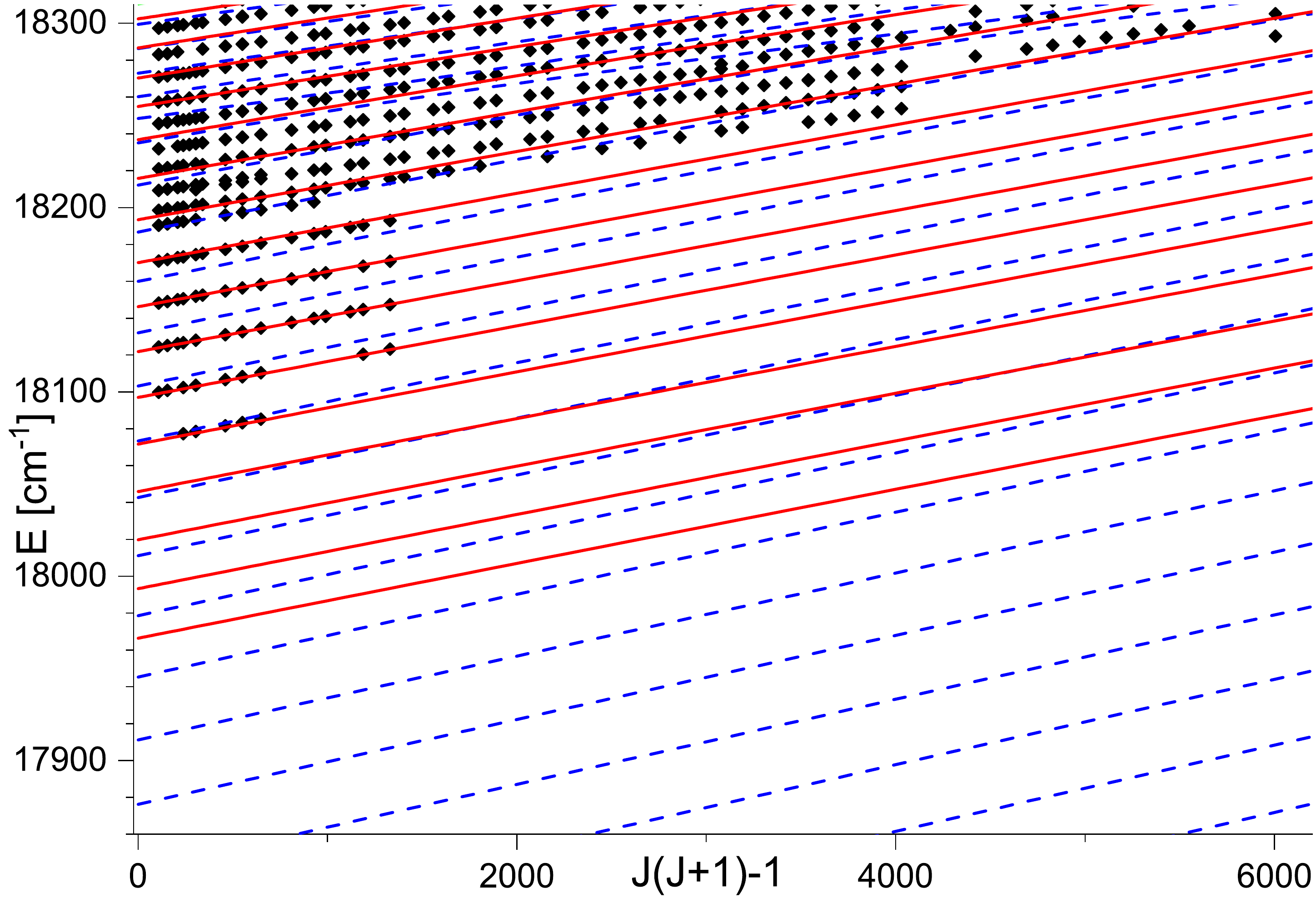}
        \label{fig:KorekC}
    \end{subfigure}
    ~
    \begin{subfigure}[t]{0.47\textwidth}
\caption{Hund's case (c)~PC}
        \includegraphics[width=\textwidth]{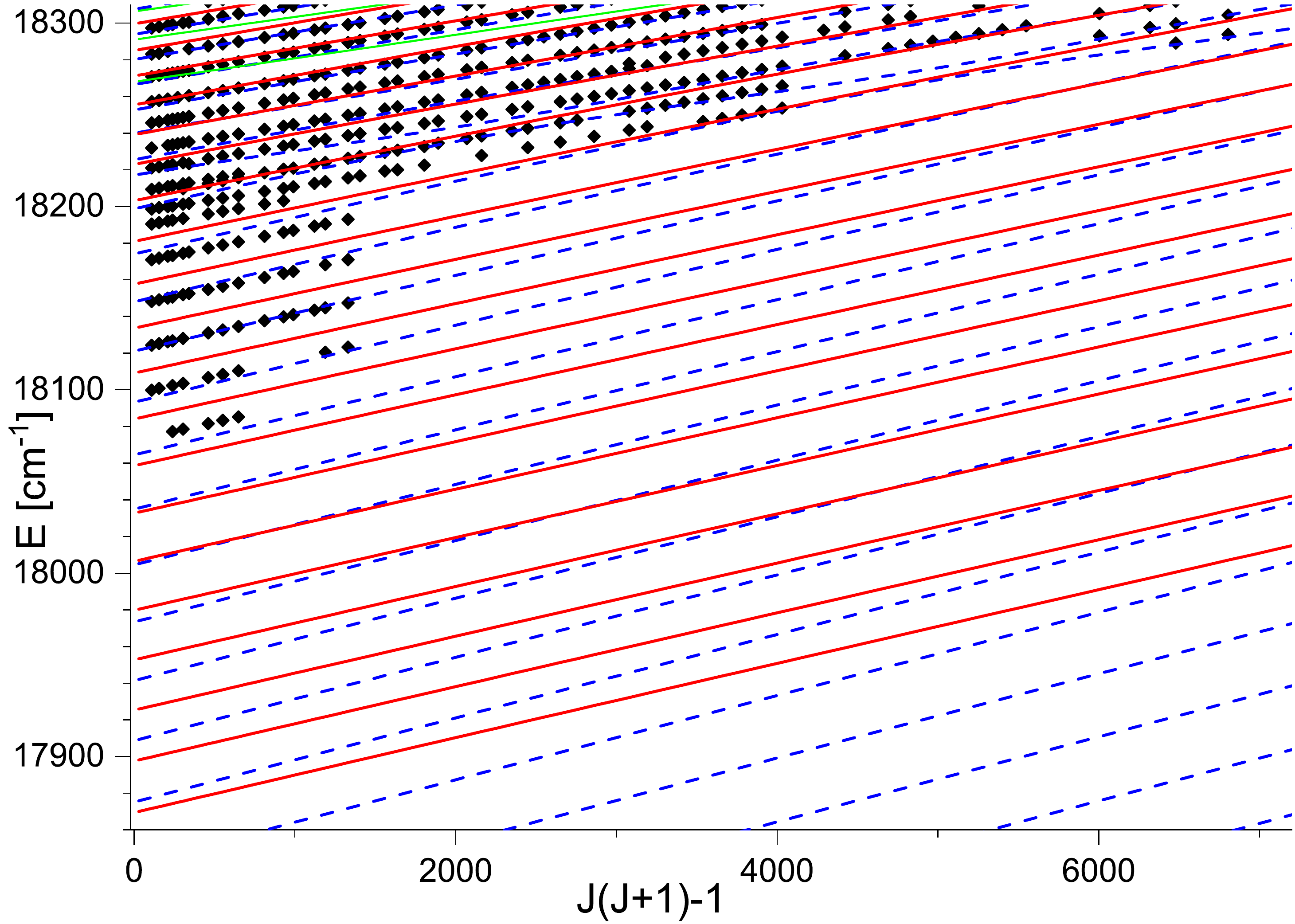}
        \label{fig:OlivierC}
    \end{subfigure}
     \caption{Energies of experimentally observed rovibrational levels (black points) compared with theoretical predictions for two Hund's coupling cases (a) and (c), corresponding to various theoretical approaches: in Fig.~\ref{fig:HabliA} that of Habli \emph{et al.} \cite{Habli2020} (only for Hund's case (a)), in Figs.~\ref{fig:KorekA} and \ref{fig:KorekC} of Korek \emph{et al.}~\cite{Korek2000,Korek2006} and in Figs.~\ref{fig:OlivierA} and \ref{fig:OlivierC} of present calculations. Predictions for either (1)$^3\Delta$ or (8)$\Omega$=1 state are marked with blue dashed lines, while for either (3)$^1\Pi$ or (9)$\Omega$=1 state are marked with red solid lines.}

     \label{fig:EnglevKorekOlivier}
\end{figure}

Since none of the pure Hund's cases is valid throughout the investigated spectral range, both ways of description were taken into account to analyse the experimental data. In the first step Hund's case (a) representation was used to find the most probable numbering of the observed vibrational levels in the (3)$^1\Pi$ state. Results of calculations performed using two different theoretical approaches  \cite{Korek2000,Habli2020}, together with our present results (Fig. \ref{fig:pot_Hund_a}) were used to infer the line assignments. In the first three blocks in Fig.\ref{fig:EnglevKorekOlivier}(a-c) the values of energies of rovibrational levels in the lowest part of experimental data are compared with theoretical ones for the three theoretical data sets. Qualitative comparison of theoretical rotational constants with the experimental ones strongly suggests that according to all three theoretical predictions the lowest observed rovibrational levels belong to the (3)$^1\Pi$ state. However the calculations in both \cite{Korek2000} and \cite{Habli2020} predict too shallow potential well of the (3)$^1\Pi$ state, which would lead to an assignment of negative vibrational quantum number to one level recorded in our spectra (Figs.~\ref{fig:KorekA} and \ref{fig:HabliA}). In contrast, relying on our present calculations the lowest level recorded in the experiment would correspond to $v'=8$ (Fig.~\ref{fig:OlivierA}), provided that the Hund's case (a) is valid in this region. 

For Hund's case (c) description only one other set of curves has been provided so far by Korek \emph{et al.}~\cite{Korek2006} (see Fig.~\ref{fig:pot_Hund_c}). Analogous comparison of experimental and theoretical location of vibrational energy levels and rotational constants done in the Hund's case (c) framework confirms that the lowest observed levels must belong to the (9)$\Omega=1$ electronic state, corresponding to the lower energy region of the (3)$^1\Pi$ state (Fig.~\ref{fig:pot_Hund_c}). In this case both theories predict deeper potential well of this state, and according to them either $v'=4$ (Korek \emph{et al.} - Fig.~\ref{fig:KorekC}) or $v'=8$ (present calculations - Fig.~\ref{fig:OlivierC}) is the lowest recorded vibrational level of the (9)$\Omega$=1 state. However we must note that the apparent good agreement between theory and experiment in Fig.~\ref{fig:KorekC} is misleading. Actually, the simulation of Franck-Condon factors performed for both theoretical potentials supports strongly the latter case, i.e. that the lowest observed vibrational level is $v'=8$, since values of these factors for transitions from $v''=0$ in the ground state (only such transitions were recorded in the experiment) to vibrational levels corresponding to $v'$<8 are close to zero according to both theoretical predictions. Additionally for both sets of Hund's (a) and (c) PECs calculated in the present paper the vibrational assignments are consistent with each other.

\begin{figure}
	\includegraphics[width=0.95\linewidth]{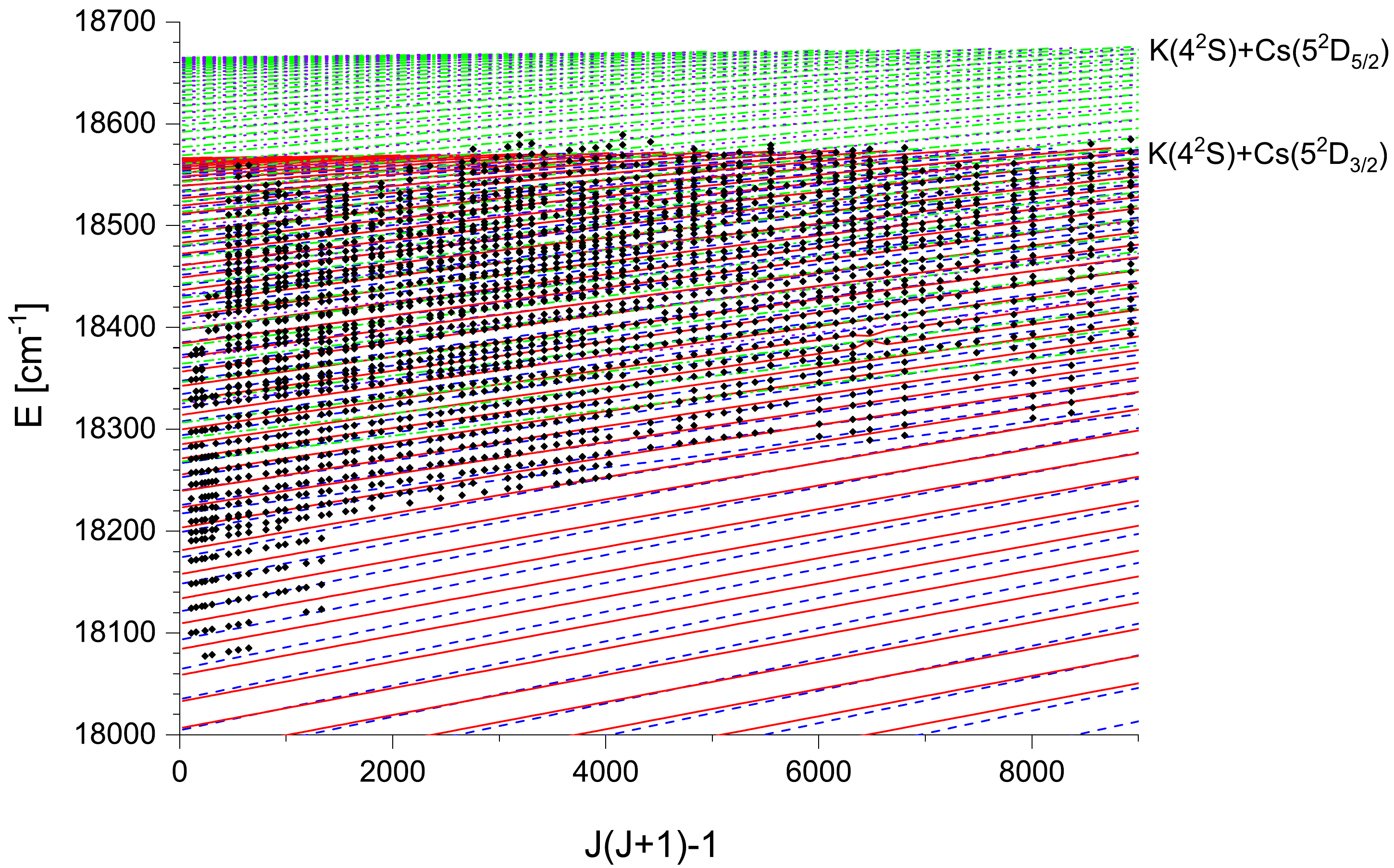}
	\caption{Energies of experimentally observed rovibrational levels (black dots) in comparison with theoretical predictions for Hund's coupling case (c) according to the calculations presented in the paper. Predictions for (8)$\Omega$=1 state are marked with dashed lines (blue in colour), for (9)$\Omega$=1 state with solid red lines, for (10)$\Omega$=1 state with green dash-dotted lines, and for (11)$\Omega$=1 state are marked with violet dotted lines.}
	\label{fig:exp_Hund_c}
\end{figure}

In Fig.~\ref{fig:exp_Hund_c} experimental energies of rovibrational levels lying close to the asymptote are compared with the theoretical ones, basing on the present calculations. Only Hund's case (c) description is valid in this region. Apparently some of the levels observed in the experiment lie above the asymptote K(4S)+Cs(5D$_{3/2})$, therefore they must belong to either (10)$\Omega$=1 or (11)$\Omega$=1 state correlated with the higher K(4S)+Cs(5D$_{5/2})$ atomic limit.

The mixture of the four (below the lower K(4$^2$S$_{1/2}$)+Cs(5$^2$D$_{3/2})$ asymptote) or two (above it) states induced by SO coupling results in strong perturbations in the molecular spectra: thus an independent description of individual states is practically impossible here. Clearly all the observed energy levels have a mixed singlet/triplet character and solely the rotational quantum number $J$ can be used to characterise a given energy level. Only a complex four-channel deperturbation procedure, would enable the determination of contribution of each of the involved states to the observed rovibrational levels. However the experimental knowledge of exact values of energies of rovibrational levels of such mixed character may be of use in planning experiments in which mixed singlet/triplet character allows for transfer of molecules to electronic states inaccessible from the ground state because of selection rules. Energies of all rovibrational levels observed in the present experiment are provided in the supplementary material which can be found, in the online version~\cite{supplC}
\section{Conclusion}

Using the well-established and powerful approach of polarisation labelling spectroscopy, we were able to record spectra for several highly-excited electronic states of the $^{39}$KCs molecule at room temperature, with a resolution of about 0.1~cm$^{-1}$ allowing for rotational resolution. Depending on the spectral range, the spectra appear as quite regular, or strongly perturbed due to efficient spin-orbit coupling between the molecular states. Guided with up-to-date theoretical calculations, we were able to assign the spectra to four different electronic states, thus confirming the quality of our electronic structure calculations. In particular, we attribute the observed perturbations to strong singlet-triplet mixing of the excited states. Further studies on these states could be undertaken using the alternative perturbation facilitated optical-optical double resonance (PFOODR) spectroscopic method \cite{Yiannopoulou1995}.

Such singlet-triplet mixing could be of relevance as an alternative scheme for the optical formation of ultracold molecules, in the case of Na$^{40}$K involving excited electronic state in a spectral range comparable to the present one, following the proposal of \cite{seesselberg2018}. Indeed, ultracold samples of KCs molecules are currently under investigation but not yet observed. Thus a possible formation scheme could rely on an initial preparation of KCs molecules in triplet-dominated magnetic Feshbach resonance \cite{Patel2014,Groebner2017}, which then could be excited at large internuclear distances to the $\Omega=1$ states composed of (3)$^3\Pi$ and (4)$^3\Sigma^{+}$ states. Both are mixed to the (3)$^1\Pi$ state, inner turning point of which lies at the same distance as the bottom of the well of the X$^1\Sigma^{+}$ electronic ground state. An adiabatic optical transfer scheme could then efficiently  transfer the population of the Feshbach resonant state towards the KCs ground state.

\bibliography{KCs3PI}

\begin{thebibliography}{10}
\expandafter\ifx\csname url\endcsname\relax
  \def\url#1{\texttt{#1}}\fi
\expandafter\ifx\csname urlprefix\endcsname\relax\def\urlprefix{URL }\fi
\expandafter\ifx\csname href\endcsname\relax
  \def\href#1#2{#2} \def\path#1{#1}\fi

\bibitem{SzczepkowskiKCs2018}
J.~Szczepkowski, A.~Grochola, P.~Kowalczyk, W.~Jastrzebski, Spectroscopic study
  of the { C(3)$^{1}\Sigma^{+}$ $\leftarrow$ X$^{1}\Sigma^{+}$ and
  c(2)$^{3}\Sigma^{+}$ $\leftarrow$ X$^{1}\Sigma^{+}$} transitions in {KCs}
  molecule, J. Quant. Spectrosc. Radiat. Transfer 204 (2018) 131--137.

\bibitem{SzczepkowskiKCs2020}
J.~Szczepkowski, A.~Grochola, P.~Kowalczyk, W.~Jastrzebski, Determination of
  the {C(3)$^{1}\Sigma^{+}$} state potential energy curve in {KCs} molecule
  based on polarisation labelling spectroscopy data, Spectrochim. Acta, Part A
  224 (2020) 117331.

\bibitem{Szczepkowski2012}
J.~Szczepkowski, A.~Grochola, W.~Jastrzebski, P.~Kowalczyk, On the
  {$4^{1}\Sigma^{+}$} state of the {KCs} molecule, J. Mol. Spectrosc. 276-277
  (2012) 19--21.

\bibitem{Szczepkowski2013}
J.~Szczepkowski, A.~Grochola, W.~Jastrzebski, P.~Kowalczyk, Study of the
  {$4^{1}\Pi$} state in {KCs} molecule by polarisation labelling spectroscopy,
  Chem. Phys. Lett. 576 (2013) 10--14.

\bibitem{Jastrzebski2015}
W.~Jastrzebski, P.~Kowalczyk, J.~Szczepkowski, A.-R. Allouche, P.~Crozet,
  A.~Ross, High-lying electronic states of the rubidium dimer—ab initio
  predictions and experimental observation of the {5$^1\Sigma^{+}_{u}$} and
  {5$^1\Pi_u$} states of {Rb$_2$} by polarization labelling spectroscopy, J.
  Chem. Phys. 143~(4) (2015) 044308.

\bibitem{Birzniece2012}
I.~Birzniece, O.~Nikolayeva, M.~Tamanis, R.~Ferber, {$B(1)^{1}\Pi$} state of
  {KCs}: High-resolution spectroscopy and description of low-lying energy
  levels, J. Chem. Phys. 136~(6) (2012) 064304.

\bibitem{Ferber199753}
R.~Ferber, W.~Jastrzebski, P.~Kowalczyk, Line intensities in {V-type}
  polarization labelling spectroscopy of diatomic molecules, J. Quant.
  Spectrosc. Radiat. Transfer 58~(1) (1997) 53--60.

\bibitem{Vexiau2017}
R.~Vexiau, D.~Borsalino, M.~Lepers, A.~Orban, M.~Aymar, O.~Dulieu,
  N.~Bouloufa-Maafa, {Dynamic dipole polarizabilities of heteronuclear alkali
  dimers: optical response, trapping and control of ultracold molecules}, Int.
  Rev. Phys. Chem. 36~(4) (2017) 709--750.

\bibitem{Borsalino2016}
D.~Borsalino, R.~Vexiau, M.~Aymar, E.~Luc-Koenig, O.~Dulieu, N.~Bouloufa-Maafa,
  Prospects for the formation of ultracold polar ground state {KCs} molecules
  via an optical process, J. Phys. B: At., Mol. Opt. Phys. 49~(5) (2016)
  055301.

\bibitem{Orban2015}
A.~Orban, R.~Vexiau, O.~Krieglsteiner, H.~C. N\"{a}gerl, O.~Dulieu,
  A.~Crubellier, N.~Bouloufa-Maafa, Model for the hyperfine structure of
  electronically excited {KCs} molecules, Phys. Rev. A 92~(3) (2015) 032510.

\bibitem{Orban_KCs_HFs2019}
A.~Orb{\'{a}}n, T.~Xie, R.~Vexiau, O.~Dulieu, N.~Bouloufa-Maafa, Hyperfine
  structure of electronically-excited states of the
  $^{39}\mathrm{K}^{133}\mathrm{Cs}$ molecule 52~(13) (2019) 135101.

\bibitem{Korek2000}
M.~Korek, A.~R. Allouche, K.~Fakhreddine, A.~Chaalan, Theoretical study of the
  electronic structure of {LiCs}, {NaCs}, and {KCs} molecules, Can. J. Phys.
  78~(11) (2000) 977--988,
  https://sites.google.com/site/allouchear/Home/diatomic.

\bibitem{Kim2009}
J.~T. Kim, Y.~Lee, A.~V. Stolyarov, Quasi-relativistic treatment of the
  low-lying {KCs} states, J. Mol. Spectrosc. 256~(1) (2009) 57--67.

\bibitem{aymar2005}
M.~Aymar, O.~Dulieu, Calculation of accurate permanent dipole moments of the
  lowest {$^{1,3}\Sigma^+$} states of heteronuclear alkali dimers using
  extended basis sets, J. Chem. Phys. 122 (2005) 204302.

\bibitem{aymar2006a}
M.~Aymar, O.~Dulieu, Comment on calculation of accurate permanent dipole
  moments of the lowest {$^{1,3}\Sigma^+$} states of heteronuclear alkali
  dimers using extended basis sets, J. Chem. Phys. 125 (2006) 047101.

\bibitem{guerout2010}
R.~Gu\'erout, M.~Aymar, O.~Dulieu, Ground state of the polar
  alkali-metal-atom--strontium molecules: Potential energy curve and permanent
  dipole moment, Phys. Rev. A 82 (2010) 042508.

\bibitem{Habli2020}
H.~Habli, L.~Mejrissi, S.~Jellali, B.~Oujia, Spectroscopic proprieties of the
  ground and the higher excited states of the {KCs}, J. Phys. B: At., Mol. Opt.
  Phys. 53~(23) (2020) 235102.

\bibitem{cimiraglia1985}
R.~Cimiraglia, J.~P. Malrieu, M.~Persico, F.~Spiegelmann, Quasi-diabatic states
  and dynamical couplings from ab initio {CI} calculations: a new proposal, J.
  Phys. B: At., Mol. Opt. Phys. 18 (1985) 3073--3084.

\bibitem{angeli1996}
C.~Angeli, M.~Persico, Quasi-diabatic and adiabatic states and potential energy
  curves for {N}a-{C}d collisions and excimer formation, Chem. Phys. 204 (1996)
  57--64.

\bibitem{NIST_ASD}
A.~Kramida, {Yu.~Ralchenko}, J.~Reader, {and NIST ASD Team}, {NIST Atomic
  Spectra Database (ver. 5.9), [Online]. Available:
  {\tt{https://physics.nist.gov/asd}} [2022, June 5]. National Institute of
  Standards and Technology, Gaithersburg, MD.} (2021).

\bibitem{Korek2006}
M.~Korek, Y.~A. Moghrabi, A.~R. Allouche, Theoretical calculation of the
  excited states of the {KCs} molecule including the spin-orbit interaction, J.
  Chem. Phys. 124~(9) (2006) 094309,
  https://sites.google.com/site/allouchear/Home/diatomic.

\bibitem{supplC}
See supplementary materials at http://dx.doi.org//... and
  http://dimer.ifpan.edu.pl.

\bibitem{Yiannopoulou1995}
A.~Yiannopoulou, K.~Urbanski, A.~Lyyra, L.~Li, B.~Ji, J.~Bahns, W.~Stwalley,
  Perturbation facilitated optical–optical double resonance spectroscopy of
  the {2$^3\Sigma^{+}_{g}$, 3$^3\Sigma^{+}_{g}$, and 4$^3\Sigma^{+}_{g}$
  Rydberg states of $^7$Li$_2$}, J. Chem. Phys. 102 (1995) 3024--3031.

\bibitem{seesselberg2018}
F.~See\ss{}elberg, N.~Buchheim, Z.-K. Lu, T.~Schneider, X.-Y. Luo, E.~Tiemann,
  I.~Bloch, C.~Gohle, Modeling the adiabatic creation of ultracold polar
  {${}^{23}{\mathrm{Na}}^{40}\mathrm{K}$} molecules, Phys. Rev. A 97 (2018)
  013405.

\bibitem{Patel2014}
H.~J. Patel, C.~L. Blackley, S.~L. Cornish, J.~M. Hutson, Feshbach resonances,
  molecular bound states, and prospects of ultracold-molecule formation in
  mixtures of ultracold {K} and {Cs}, Phys. Rev. A 90 (2014) 032716.

\bibitem{Groebner2017}
M.~Gr\"obner, P.~Weinmann, E.~Kirilov, H.-C. N\"agerl, P.~S. Julienne, C.~R.
  Le~Sueur, J.~M. Hutson, Observation of interspecies {Feshbach} resonances in
  an ultracold $^{39}\mathrm{K}\ensuremath{-}^{133}\mathrm{Cs}$ mixture and
  refinement of interaction potentials, Phys. Rev. A 95 (2017) 022715.

\end{thebibliography}

\end{document}